# Temperature Dependent Energy Gain of Bifacial PV Farms: A Global Perspective

M. Tahir Patel[1], Ramachandran A. Vijayan[2], Reza Asadpour[1],
M. Varadharajaperumal[2], M. Ryyan Khan[3], and Muhammad A. Alam[1]

[1]Electrical and Computer Engineering Department, Purdue University, West Lafayette, IN, USA
[2]Department of Electrical and Electronic Engineering, Sastra University, Thanjavur, TN, India
[3]Department of Electrical and Electronic Engineering, East West University, Dhaka, Bangladesh

*Abstract* – **Bifacial solar panels are perceived to be the technology of choice for next generation solar farms for their increased energy yield at marginally increased cost. As the bifacial farms proliferate around the world, it is important to investigate the role of temperature-dependent energy-yield and levelized cost of energy (LCOE) of bifacial solar farms relative to monofacial farms, stand-alone bifacial modules, and various competing bifacial technologies. In this work, we integrate irradiance and light collection models with experimentally validated, physics-based temperature-dependent efficiency models to compare the energy yield and LCOE reduction of various bifacial technologies across the world. We find that temperature-dependent efficiency changes the energy yield and LCOE by approximately −10 to 15%. Indeed, the results differ significantly depending on the location of the farm (which defines the illumination and ambient temperature), elevation of the module (increases incident energy), as well as the temperature-coefficients of various bifacial technologies. The analysis presented in this paper will allow us to realistically assess location-specific relative advantage and economic viability of the next generation bifacial solar farms.**

## 1. INTRODUCTION

The PV industry is actively developing bifacial module technology to reduce the levelized cost of energy (LCOE) of the utility-scale solar PV farms [1–12]. This is because a bifacial solar panel collects light at both the front and rear surfaces as compared to the monofacial panels that collect only at the front. A recent literature review by Guerrero-Lemus et al. [13] explains how bifacial technology has developed from its infancy in 1960s to practical applications and scalability. Despite the possibility of higher output from the same footprint, bifacial PV technology had to wait until recent innovations for an accelerated deployment. The International Technology Roadmap for Photovoltaic (ITRPV) predicts that the worldwide market share for bifacial technology will increase from 15% in 2020 to 40% by 2028 [14]. Since bifacial solar PV farm deployments are now increasing, it is important to understand how module-level bifacial gain translates to location and technology-specific energy yield and LCOE of bifacial farms. Specifically, a realistic assessment of the relative merits of emerging bifacial PV technologies must account for efficiency losses from thermal effects of the bifacial solar panels installed in utility-scale solar farms. Other spectral and angle-of-incidence effects also affect efficiency during the early morning and late evening hours of the day. However, these effects are diminished during the operating hours of the PV farms which contribute to the energy production.

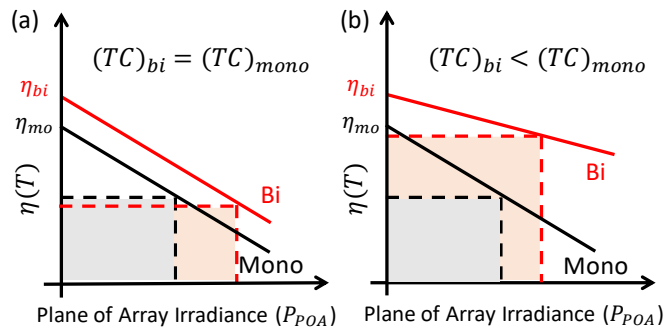

Fig. 1 Effect of temperature coefficient ($TC$) and irradiance on efficiency of monofacial (black) and bifacial (red) solar panels. Note that energy yield is a product of Plane of Array irradiance ($P_{POA}$) and efficiency. Hence, higher energy yield requires higher efficiency and higher $P_{POA}$. (a) Same PV technology; (b) different PV technologies.

Fig. 1 shows a linear relation between efficiency and irradiance (for typical operating conditions) and summarizes the importance of temperature-aware performance modeling of solar cells. The output power ($P_{out}$) is determined by the temperature-dependent efficiency ($\eta(T)$) and the input irradiance ($P_{POA}$), i.e., $P_{out} = \eta(T)P_{POA}$. Therefore, various technologies can be compared simply by comparing the rectangular boxes in Fig. 1. Since the cell temperature is a function of the ambient temperature $T_{amb}$, input power $P_{POA}$, environmental conditions (such as wind speed, relative humidity, and sky temperature), mounting configuration and module construction, and technology-specific thermal coefficient of maximum power, $TC$, i.e., $T_{cell} = f(T_{amb}, P_{POA}, TC, ...)$, the conclusions reached regarding bifacial gain based on temperature-independent efficiency, $\eta_{STC}$, [15] may not translate in practice. Indeed, the bifacial gain may be negative or positive depending on the geographical location that determines $T_{amb}$ and $P_{POA}$ as well as PV technology which affects the $TC$ and the efficiency gain ($\eta_g = (\eta_{bi}/\eta_{mo})_{STC}$), where $\eta_{mo}$ is the efficiency of the monofacial cells while $\eta_{bi}$ is the normalized output of bifacial cells [16,17]. If the bifacial and monofacial panels have the same temperature coefficients (i.e. $(TC)_{bi} = (TC)_{mono}$, as in Fig. 1(a)), higher irradiance would lead to additional heating of bifacial panels, leading to higher degradation of efficiency. Thus, bifacial panels would outperform monofacial panels only when $P_{POA} < (2\eta_g - 1)/(k(TC)(4\eta_g - 1))$ where $k$ is a location-specific constant and $\eta_g$ is the bifacial-to-monofacial efficiency gain (for details, see Appendix S1). On the other hand, when bifacial panels have a lower $TC$ as compared to the monofacial panels (Fig.1(b)), even higher irradiance would not degrade the efficiency below that of monofacial panels. Thus, bifacial PV panels have an optical advantage over monofacial panels but thermal loss at higher temperatures of operation may potentially negate some of the optical gains. In short, temperature-dependence fundamentally alters the conclusions regarding location-specific bifacial energy gain across the world.

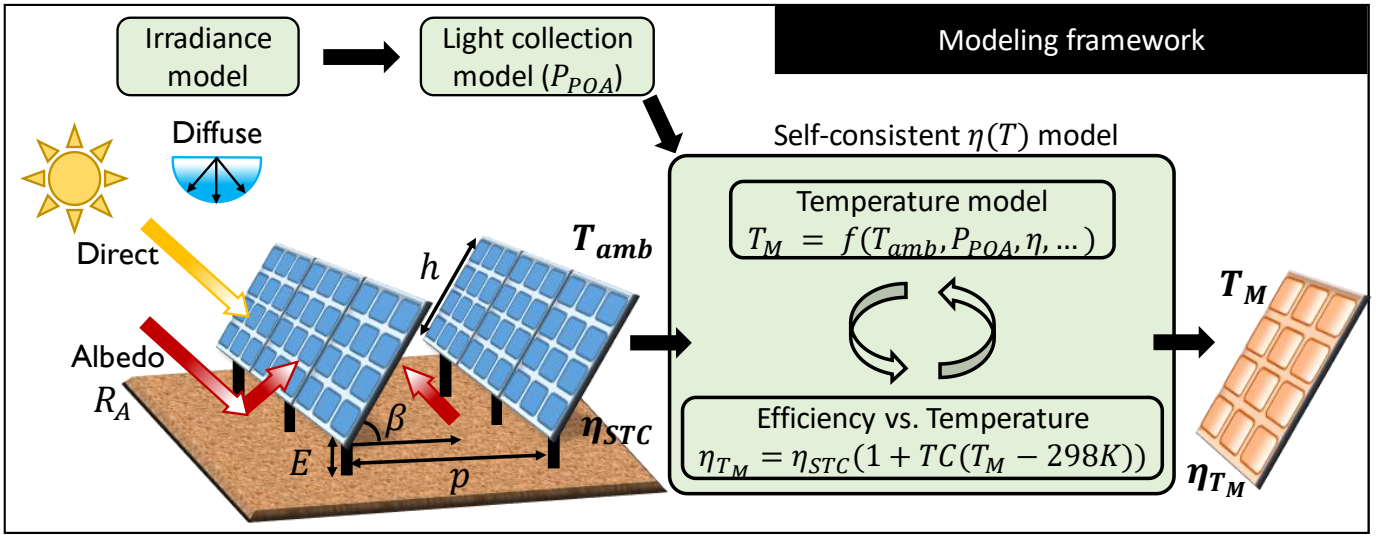

Fig. 2 Flow diagram of the modeling framework with physical parameters for temperature-dependent energy yield estimation.

Recent studies have therefore focused on the temperature-dependent performance of *stand-alone* monofacial and bifacial PV modules [18–23]. For example, Ref. [18–20] contains a systematic and detailed analysis of temperature coefficients for different PV materials and technologies. The open rear surfaces of a bifacial cell increase power-input due to albedo and power-loss due to IR radiation. In principle, therefore, the bifacial cells can operate cooler than a monofacial cell [22]. A number of theoretical studies have explored the temperature-dependence of several performance parameters of solar cells as well as solar PV systems [21,23]. Lopez-Garcia et al. have experimentally (indoor and outdoor) analyzed the temperature coefficients ($TC$) of bifacial c-Si PV modules. They concluded that bifacial $TC$ was observed to be lower than monofacial $TC$ [24]. Recent studies [25–28] based on coupled optical-electrical-thermal model of the bifacial photovoltaic module were performed that observed a 22% bifacial gain in yearly energy yield and suggested tilt and elevation optimizations for stand-alone bifacial module performance. In short, many groups have carefully studied the temperature-dependent performance of *stand-alone* modules.

Compared to the extensive literature for stand-alone modules, there are only a few reports of temperature-independent yield loss of various monofacial and bifacial technologies at the farm level [15,29,30]. The standalone module and a farm differ significantly in terms of albedo collection related to the row-to-row mutual shading, leading to substantial changes in self-heating. In this paper, we model a generalized bifacial solar PV farm that includes the effects of temperature-dependent efficiency, as portrayed in Fig. 2.

The integrated model will combine optical, thermal, and electrical models with economic analysis to estimate the energy yield and LCOE of a solar PV farm with focus on bifacial modules, see Fig. 2. The effects of temperature on a module and farm at different locations over the globe are compared and it is found that the *stand-alone* modules are affected by temperature more than the farms. Moreover, the roles of temperature coefficients of different materials (e.g., Al-BSF and SHJ) are explored from the perspectives of efficiency degradation and bifacial gain (bifacial vs. monofacial). Lower temperature coefficient ($0.26\%/°C$) for SHJ (bifacial) shows reduced efficiency degradation and yields higher bifacial gain (**20-40% globally**) compared to Al-BSF (monofacial) which has a higher temperature coefficient ($0.41\%/°C$). Next, combined dependencies of temperature and elevation on the energy yield are considered and it is observed that the energy yield saturates after a threshold value of elevation (typically, $E_0 = 2m$). About **$1-15$%** increase in energy yield between elevated ($E = 2m$) and unelevated ($E = 0m$) bifacial farms is observed around the world.

In Sec. 2 we describe the optical, thermal, and electrical models used in our study and the associated equations are organized in Tables I and II. In Sec. 3 we show the results and discuss important trends and features. We vary several design parameters of a solar PV farm e.g., elevation, PV technology, bifacial/monofacial, and $TC$, to quantify the effects of temperature-dependent efficiency on energy yield, LCOE, and optimum design (tilt angle) of the farm. Finally, in Sec. 4, we summarize the study and conclude the paper.

## 2. MODELING AND VALIDATION

An LCOE-aware modeling of bifacial solar farm involves calculation of irradiance, collection of sunlight and electrical output coupled to economic analysis. The optimized design (panel orientation, tilt, elevation, and spacing) of the farm then can be evaluated in terms of energy or LCOE.

Consider a solar farm oriented at an azimuth angle $\gamma_A$ from the North, with panels of height $h$, tilted at an angle $\beta$, elevated from the ground to $E$, and with pitch $p$ over a ground of albedo $R_A$. In the following discussions, we will focus on three aspects of the modeling framework for the specified solar farm: (A) Re-parameterizing LCOE, (B) Irradiance modeling, and (C) Collection of light. These topics are discussed below.

TABLE I
EQUATIONS ASSOCIATED WITH OPTICAL MODEL: IRRADIANCE MODELING AND LIGHT COLLECTION

| | |
|---|---|
| (1) $I_{GHI} = I_{DNI} \times \cos(\theta_Z) + I_{DHI}$ | (6) $I_{PV:Alb.diff}^{F,Panel}(l) = I_{Gnd:DHI} \times R_A \times F_{dl-gnd}(E,l) \times \eta_{diff}$ |
| (2) $I_{PV:DNI}^{F,Farm} = I_{DNI} \cos\theta_F (1 - R(\theta_F)) \eta_{dir}$ | (7) $I_{PV:Alb.diff}^{F,Farm} = 1/h \int_0^h I_{PV:Alb.diff}^{F,Panel}(E,l) \, dl$ |
| (3) $I_{PV:DHI}^{F,Farm} = I_{DHI} \, \eta_{diff}/h \int_0^h \frac{1}{2}(1 + \cos(\psi(E,l) + \beta)) \, dl$ | (8) $I_{PV:Alb}^{Farm} = I_{PV:Alb.dir}^{Farm} + I_{PV:Alb.diff}^{Farm}$ |
| (4) $I_{PV:Alb.dir}^{F,Panel}(l) = I_{Gnd:DNI} \times R_A \times F_{dl-gnd}(E,l) \times \eta_{diff}$ | (9) $I_{PV(T)}^{Total} = I_{PV:DNI}^{Farm} + I_{PV:DHI}^{Farm} + I_{PV:Alb}^{Farm}$ |
| (5) $I_{PV:Alb.dir}^{F,Farm}(l) = 1/h \int_0^h I_{PV:Alb.dir}^{F,Panel}(E,l) \, dl$ | (10) $YY_T(p, \beta, h, E, \gamma_A, R_A) = \int_0^1 I_{PV(T)}^{Total}(p, h, E, \gamma_A, R_A) \, dY$ |

TABLE II
EQUATIONS ASSOCIATED WITH THERMAL MODEL

| | |
|---|---|
| (11) $\eta(T_{cell}) = \eta_{STC}(1 - TC(T_{cell} - T_{STC}))$ | (14) $T_{cell} = T_{amb} + (P_{POA} - \eta P_{POA} - \gamma P_{POA}(subband)) * \frac{F}{1000}$ |
| (12) $T_{cell} = T_{amb} + P_{POA} \times e^{a+(b \times WS)} + \Delta T \times \frac{P_{POA}}{1000}$ | (15) $T_{cell} = T_{amb} + \left( P_{POA_{top}} - \eta_{top} \times P_{POA_{top}} + P_{POA_{bot}} - \eta_{bot} \times P_{POA_{bot}} - P_{POA}(subband) \right) * \frac{F}{1000}$ |
| (13) $T_{cell} = T_{amb} + c_T \frac{\tau\alpha}{u_t} P_{POA} \left( 1 - \frac{\eta(T_{cell})}{\tau\alpha} \right)$ | (16) $P_{POA} = I_{PV:DNI}^{Farm}/\eta_{dir} + (I_{PV:DHI}^{Farm} + I_{PV:Alb}^{Farm})/\eta_{diff}$ |

## 2.1 Farm Model with LCOE*

LCOE is defined as the ratio of the total cost of a PV system to the total energy yield of the system over its lifetime [31], i.e.,

$$LCOE = \frac{\text{Total Cost (\$)}}{\text{Total Energy Yield (kWh)}}$$

$$= \frac{C_{sys}(Y=0) + (\sum_{k=1}^{Y} C_{om}(k)) - C_{rv}(Y)}{E(Y)}$$

(1)

where $C_{sys}(Y = 0)$ is the initial fixed installation cost of the system (i.e., at $Y = 0$). $C_{sys}(Y = 0)$ includes the cost of modules ($c_{m,0}$), the cost of land ($c_{l,0}$), and the balance of system cost ($c_{bos,0}$) such as labor, permit, racks, inverters, etc. The recurring operations and maintenance cost ($C_{om}$) scales with the cost of maintaining individual modules ($c_{om,m}$) and the cost of maintaining the land ($c_{om,l}$). Finally, $C_{rv}$ is the residual value of the modules ($c_{rv,m}$), the land ($c_{rv,l}$), and the equipment to be regained when the farm is decommissioned. $C_{om}$ and $C_{rv}$ are a function of the lifetime (number of years, $Y$) for which the solar farm is operated.

Since the costs vary with the number and size of PV modules and the solar-farm land, the LCOE expression can be described using the dimensions of a solar farm and the modules installed, as shown in Eq. (2) below.

$$LCOE = \frac{\mathbb{C}_M(r).h.M.Z + \mathbb{C}_L(r).p.M.Z + C_{bos,f}}{YY_T(p, h, E, \beta, \gamma_A, R_A).M.Z.h.\chi(d,r)}$$

(2)

Here $\mathbb{C}_M$ is the cost per unit meter of module (height), $\mathbb{C}_L$ is the cost per unit meter of land (pitch), $M$ is the number of rows/arrays of modules and $Z$ is the number of modules in an array (in the z-direction, into the page). $YY(= E_0)$ is the yearly energy yield per meter of a pristine module for one period/pitch ($p$) such that the yearly energy of the farm is $E(Y) = YY_T.M.Z.h.\chi(d,r)$, where $\chi = \sum_{k=1}^{Y}(1-d)^k(1+r)^{-k}$. $YY_T$ is a function of the physical design parameters $(p, h, E, \beta, \gamma_A, R_A, T)$. The lifetime ($Y$, typically 25 years) of a solar farm is defined as the time duration before the performance (efficiency) of a solar farm degrades by 20%. Thus, the degradation rate ($d$, typically 0.7%/year) defines the lifetime of a solar farm. The discount rate ($r$) accounts for the devaluation of predicted future earnings.

Further, the cost associated with the balance of system ($c_{bos,f}$) is typically negligible as compared to the essential costs, $C_M$ and $C_L$ [12,32], and does not affect the design optimization of the farm. With these considerations, we arrive at the 'essential levelized cost of energy' ($LCOE^*$), as follows:

$$LCOE^* = \frac{LCOE. \chi}{\mathbb{C}_L} = \frac{\mathbb{C}_M/\mathbb{C}_L + p/h}{YY_T} = \frac{p/h + M_L}{YY_T}$$

(3)

Here, $M_L(\equiv \mathbb{C}_M/\mathbb{C}_L)$ is the ratio of the cost of module per unit length (height) to the cost of land per unit length (pitch). $M_L$ essentially captures the costs of a solar farm whereas $p/h$ and $YY_T(p, h, E, \beta, \gamma, R_A)$ contains the information about the physical parameters of the farm. $M_L$ varies with the technology and location in the world. In this paper, we assume a typical value of $M_L = 15$ [15]. Given $M_L$, we can perform the physical design optimization of a solar farm to find a minimum $LCOE^*$. Equation (3) shows that $LCOE$ is proportional to $LCOE^*$, since $\mathbb{C}_L$ and $\chi$ are location-specific constants. Therefore, minimizing $LCOE$ is equivalent to minimizing $LCOE^*$ for a given location. Ref. [15] provides a more detailed derivation and discussion on $LCOE$ and $LCOE^*$.

Next, we will estimate the amount of sunlight falling on a farm at any location in the world.

## 2.2 Irradiance Modeling and Light Collection

The energy yield of a bifacial solar farm for a given albedo is numerically modeled in three steps.

*(1) Irradiance model.* we calculate the amount of sunlight incident at a location defined by its latitude and longitude. This requires the Sun's trajectory (zenith ($\theta_Z$) and azimuth angle ($A$)) and the irradiance [29]. The Global Horizontal Irradiance (GHI or $I_{GHI}$) is ideally given by Haurwitz clear sky model [33,34]. We renormalize this irradiance based on the NASA Surface meterology and Solar Energy database [35] to find the local variation in GHI. The GHI is split into direct light (DNI or $I_{DNI}$) and diffuse light (DHI or $I_{DHI}$) using Orgill and Hollands model [36].

*(2) Light collection on panels.* we quantify the amount of light collected by the solar panels installed at that location. The panels have height $h$, tilted at an angle $\beta$, separated by pitch (or period) $p$, and are oriented at array azimuth angle $\gamma_A = 180°$ (i.e., south-facing panels) for farms in the northern hemisphere and $\gamma_A = 0°$ (i.e., north-facing panels) for farms in the southern hemisphere or $\gamma_A = 90°$ for panels facing E-W direction. The collection of light on panels from the three components of irradiance, i.e., direct, diffuse, and albedo light are formulated separately and analyzed accordingly. Our approach to model the collection of light follows Ref. [15] and the equations are summarized in Table 1. We calculate the angle of incidence (AOI) of light to find the component of direct light ($I_{DNI}$) falling on the tilted panel's front and/or the back face (depending on the tilt angle). $\theta_F$ is AOI for front face, $R$ is the angle-dependent reflectivity ($R(\theta_F)$) of the panel, $\psi(l)$ is the viewing angle at length $l$, and $F_{dl-gnd}$ is the view factor. We extend the ground mounted panel-array model in [15] to include calculations for elevated panels. The pattern of light falling on the ground, estimated by the view factor ($F_{gnd-sky}$), and subsequently the albedo light collected by the panels, estimated by the view factor ($F_{dl-gnd}$), both vary with the elevation (E) of the panels. A detailed explanation of the albedo light varying with elevation is presented in the appendix.

*(3) Energy yield.* We finally find the daily and yearly energy-output of the farm. Using the equations (2), (3), and (8) in Table 1, we arrive at Eq. (9) to find the time-varying spatially distributed light collection on the panels. This information is used in the circuit model to find the equivalent power generation. To estimate energy output, we integrate the power generated over the desired period of time. We define the energy yield per pitch of a farm over one year as yearly yield (*YY*) given by Eq. (10). Now, we can vary the albedo to determine the variation in *YY*.

## 2.3 Temperature-dependent efficiency models

The temperature-dependent efficiency loss involves a complex interplay of increased absorption due to bandgap reduction vs. increased dark-current and reduced mobility. At practical illumination intensity, it is well-known that the efficiency $\eta(T)$ scales linearly with the module/cell temperature, see Eq. 1 in Table 1 where the rate/slope of degradation is given by the absolute temperature coefficient (*TC*) [21,37].

The module/cell temperature further depends on the ambient temperature ($T_{amb}$), wind speed ($W_S$), module mounting type,

efficiency of the module, and other factors. In this work, we have considered King (Eq. 12, Table II), Skoplaki (Eq. 13, Table II) and Alam-Sun (Eq. 14, Table II) models to estimate the module/cell temperature based on location-specific the ambient temperature.

Briefly, **King's Model** (Eq. 12) accounts for the wind speed (WS) and cell to module temperature differences ($\Delta T$) where the parameters 'a' and 'b' are negative and are used as fitting parameters for the module type and deployment. Notice that this simple model does not explicitly account for output power (efficiency is absent in the analytical expression) hence, it doesn't need to be solved self-consistently with Eq. 11, Table II.

Unlike King's model, **Skoplaki's model** (Eq. 13) model accounts for the module efficiency. It also includes:(i) $u_L$ which is the heat loss coefficient used as a fitting parameter to the experimental energy yield and found to be 21.5 for our experimental data (ii) $c_T$ is the correction term which accounts for averaging the ambient temperature and irradiance. We found that instantaneous data (every 1-min) does not require any correction term to predict the energy yield. In addition, we found that $c_T$ increases with the duration of the average i.e. daily average requires smaller correction term than monthly average.

Finally, the **Alam-Sun** (Eq. 14) model generalizes the Skoplaki's model, which allows us to account for the temperature rise due to the sub-band gap absorption. The model uses a parameter ($0 < \gamma < 1$) to account for the fraction of sub-bandgap irradiance/power that is not absorbed by the solar module. When $\gamma = 0$, Sun-Alam model reduces to Skoplaki model. Sub-bandgap absorption is an important consideration because the bifacial glass-glass modules may have reduced sub-bandgap absorption compared to monofacial glass-backsheet modules. The relatively cool bifacial modules (for the same $P_{POA}$) would improve the bifacial gain. In fact, a slightly generalized version of Alam-Sun model is necessary to account for the asymmetric absorption from the rear side of the module.

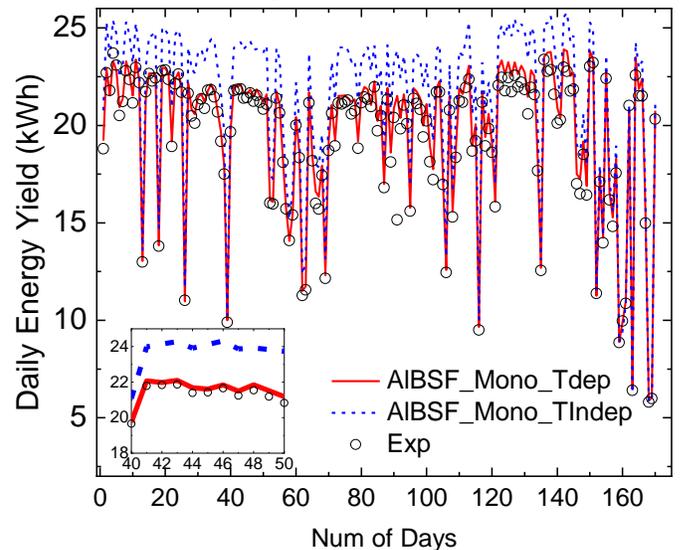

Fig. 3 Daily and Monthly energy yield for Albuquerque, NM, USA. Simulation is done based on the nameplate efficiency along with and without its temperature dependence. (Inset) A zoomed-in version of Fig. 3 from day 40 to 50. Please note that we have not considered any other sophisticated models like spectral irradiance here.

These cell temperature estimation models are coupled to the temperature dependence of efficiency and solved self-consistently. Once the temperature-dependent efficiency is obtained, we multiply it with the incident light intensity to deliver the energy output (i.e. the rectangular box in Fig. 1). Figs. 4 demonstrates the validity of the $T$-dependent simulation model with the experimental data from Sandia National Labs, Albuquerque, NM, USA. We use the experimentally available $P_{in}$ along with the Skoplaki model to find that the experimentally observed energy yield (Black circles) accurately matches the simulated $T$-dependent energy yield (Red-solid line). Note that the $T$-independent efficiency model overestimates the energy yield (Blue-dashed line). Fig. 3(inset) is a zoomed in version of Fig. 3 which shows the accuracy of simulated T-dependent energy yield. Further, Fig. 4 summarizes the importance of the sub-bandgap absorption (Alam-Sun model, Eq. 14) and the effect of $0 < \gamma < 1$ on efficiency.

### 2.4 Energy Output with temperature dependence

Finally, the power generated by the panels under STC (described in section 2.2) is corrected using the temperature dependent efficiency model i.e., Eqs. (11), (13), and (16) from Table II. For energy output, we integrate the power generated over the desired period of time. We define the energy yield per pitch of a farm over a period of one year as yearly yield ($YY$).

$$YY_T(p, \beta, h, E, \gamma, R_A) = \int_0^1 I_{PV(T)}^{Total}(p, \beta, h, E, \gamma_A, R_A) \ dY \quad (4)$$

Fig. 3 validates our temperature-dependent energy yield prediction with experimental data from Sandia National Laboratory. The deployed panels were manufactured by Canadian Solar (CS6K-275M 275W mono-Si) with a module efficiency $\eta_{STC} = 16.8\%$, temperature coefficient $TC = 0.41\%/°C$, and outer dimensions of 1650 x 992 mm. The modules are installed S-facing, 35° fixed-tilt at Sandia National Laboratories in Albuquerque, New Mexico USA (35.05° N, 106.54° W) at an elevation of 1663 m above sea level and are mounted in a 2-up landscape fashion. Row spacing is 4.88 m. They are grid-connected in four strings of 12 modules. Performance data presented in this paper is derived from string-level I-V curves that were measured at 30-minute intervals over a period of ~170 days. The $T$-dependent energy estimation (dashed blue line) follows the experimental data (black circles) almost exactly while the $T$-independent energy (solid red line) is overestimated.

### 3. RESULTS AND DISCUSSION

As the first step of calculating the temperature-dependent energy yield and LCOE, we first examine the temperature dependence of monofacial and bifacial module configurations. The cell temperature is calculated self-consistently with efficiency and irradiance. Fig. 4 shows the trend in $\eta(T)$ between monofacial and bifacial modules. Although the bifacial modules collect more light (irradiance), the apparent temperature coefficient (slope of the red line in Fig. 4(b)) is relatively lower. This can be explained using the effect of spectral irradiance on the modules, as shown in Fig. 4(b). Referring to the Sun-Alam model, $\gamma$ is the fraction of sub-bandgap irradiance/power that is not absorbed by the solar module. Looking at the two extreme cases of $\gamma = 0$ (monofacial with opaque back sheet) and $\gamma = 1$ (bifacial with transparent back sheet), we see that the efficiency vs irradiance plot has a steeper slope for monofacial case. The rapid decrease in efficiency is a consequence of increased temperature (Fig. 4(a)) due to additional sub-bandgap (or IR) absorption.

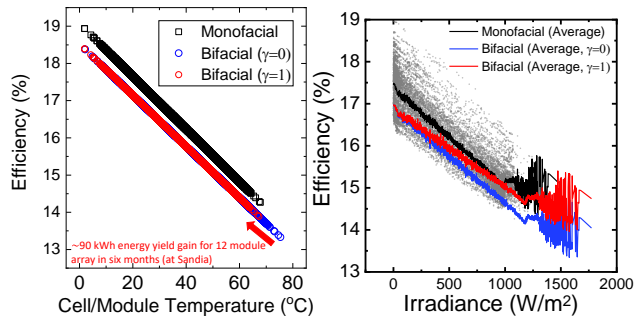

Fig. 4: (a) Efficiency vs T for both monofacial and bifacial solar panel. The $TC$ is same for both the panels which is reflected in the slopes – however, bifacial is cooler (shown as blue arrow) and lead to ~90 kWh gain when the sub-band spectra is assumed transparent ($\gamma = 1$). (b) Average efficiency vs Irradiance for both monofacial and bifacial solar panel where the background symbols (small open circles) represent the actual data for monofacial string. The change in the slope indicates the change in the temperature of the panel due to the sub-band transparency.

Since $TC$ is a significant parameter in temperature-dependent efficiency estimation, it requires an effort from the solar PV community to reduce $TC$ through material and process engineering.

### 3.1 Self-heating affects a stand-alone module more than modules in a farm

A bifacial module in a solar PV farm collects less irradiance ($P_{POA}$) as compared to a standalone bifacial module due to

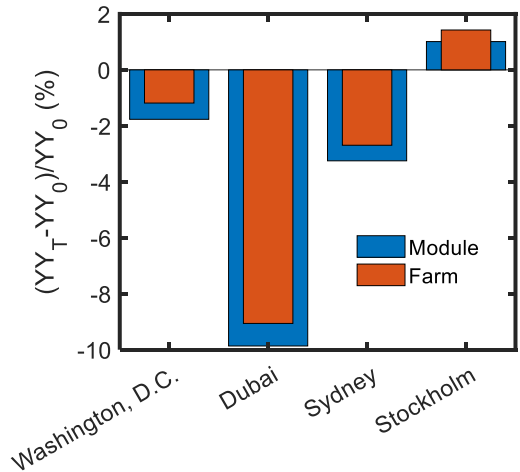

Fig. 5: Comparing module and farm-level $\eta_T$ vs. $\eta_0$ based percentage change in energy yield for different locations around the world. We observe that degradation is higher for an elevated (E=2m) standalone module as compared to an elevated farm, except for locations at very high latitude e.g., Stockholm.

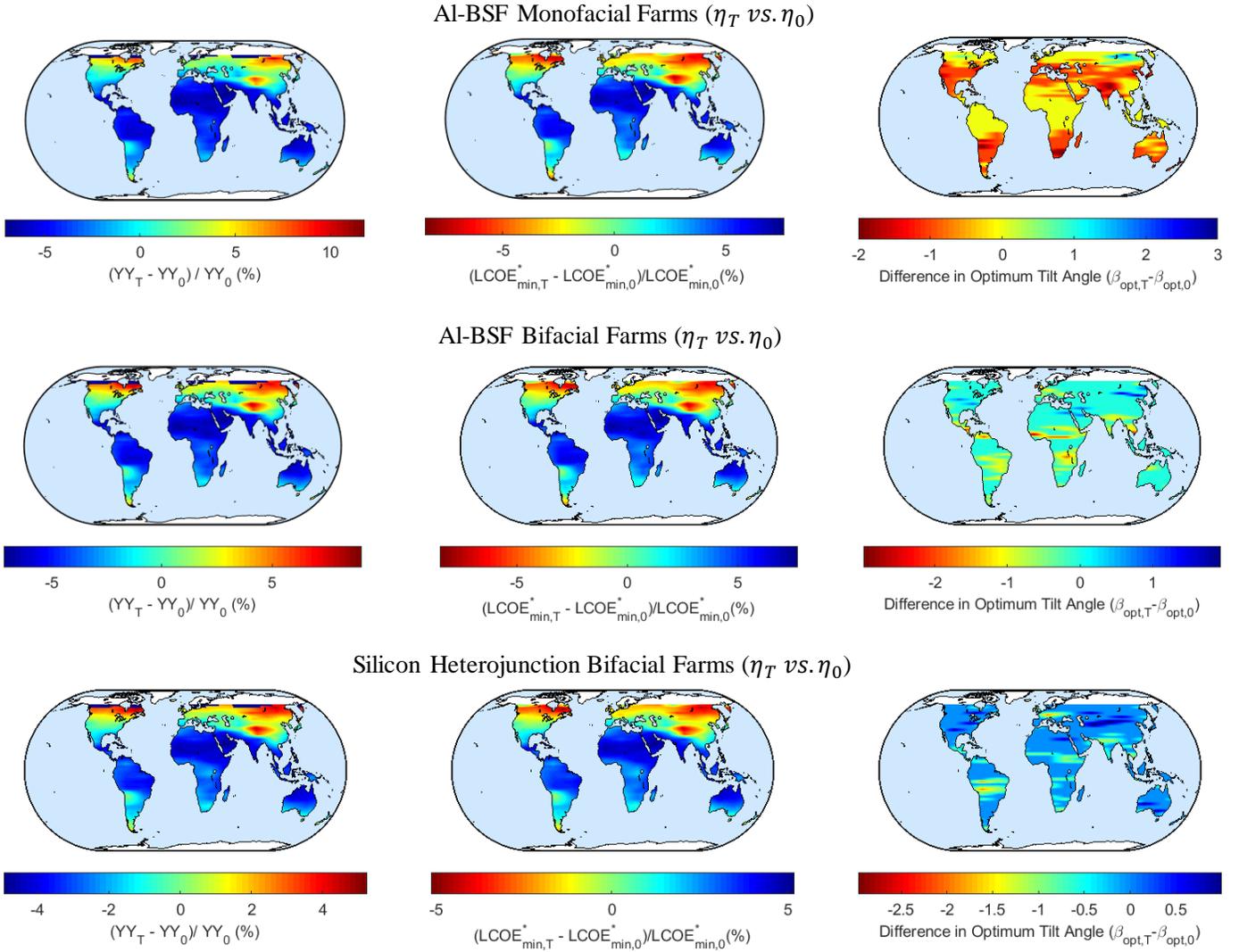

Fig. 6: Global maps comparing yearly energy yield, $LCOE^*$, and difference in optimum array tilt angle for temperature-dependent efficiency (subscript T) and temperature-independent efficiency (subscript 0). Note that $R_A = 0.5$ and $M_L = 15$. (Row 1) Monofacial Farms, e.g., Al-BSF, (Row 2) Bifacial farms with a high temperature coefficient ($TC = 0.41\%/°C$), and (Row 3) Bifacial farms with a lower temperature coefficient ($TC = 0.26\%/°C$) e.g., SHJ solar cells.

mutual shading (periodic shading on the ground). Thus, our self-consistent model predicts the temperature of a standalone module to be somewhat higher than that of a module in a farm. Note that the temperature difference would change somewhat (< 3-4 degrees) once the "heat-island effect" is accounted for [38]. The relatively small correction reflects two counter-balancing effects of ground illumination vs. heat trapping in a farm vs. in a stand-alone module. In general, the efficiency ($\eta_T$) and energy yield ($YY_T$) degradation, with respect to $\eta_{STC}(or \eta_0)$ and $YY_0$, are more prominent for a standalone module as compared to a farm, see Fig. 5. We observe that the degradation in $YY$ is enhanced for standalone bifacial modules as compared to bifacial farms. For locations with hotter climates the percentage change in $YY_T$ and $YY_0$ is negative, due to efficiency degradation at higher temperatures. On the other hand, for places with very cold climates, like Stockholm, the associated percentage change is positive and higher for a farm in comparison to a standalone module. The apparent exception at

Stockholm arises due to colder weather i.e., $T_{amb} < T_{STC}$, during most of the year. Negative ambient temperature and low irradiance ($P_{POA}$) for a farm, leads to lower $T_{cell}(farm)$ as compared to $T_{cell}(module)$, see Eq. (13). Thus, from Eq. (11), $\eta_T(farm) > \eta_T(module)$ and $\Delta YY = P_{POA} \times \Delta\eta_T$ is higher for a farm as compared to a *stand-alone* module.

### 3.2 Technology and geography dictate the temperature-dependent yield- loss of monofacial vs. bifacial farms

The worldwide maps in Fig. 6 depict the dependence of gain/loss in energy yield between temperature-dependent efficiency ($\eta_T$) vs. constant efficiency ($\eta_0$ or $\eta_{STC}$). Clearly, hotter places like the Sahara Desert, Mexico, and southern India, show a loss in energy yield due to degradation in efficiency with higher temperature as compared to efficiency at STC ($\eta_{STC}$). On the other hand, for colder locations like Siberia, Gobi Desert, and northern Canada and Europe, with lower

(Al-BSF (Bifacial) − Al-BSF (Monofacial)) / Al-BSF (Monofacial) for $\eta(T)$

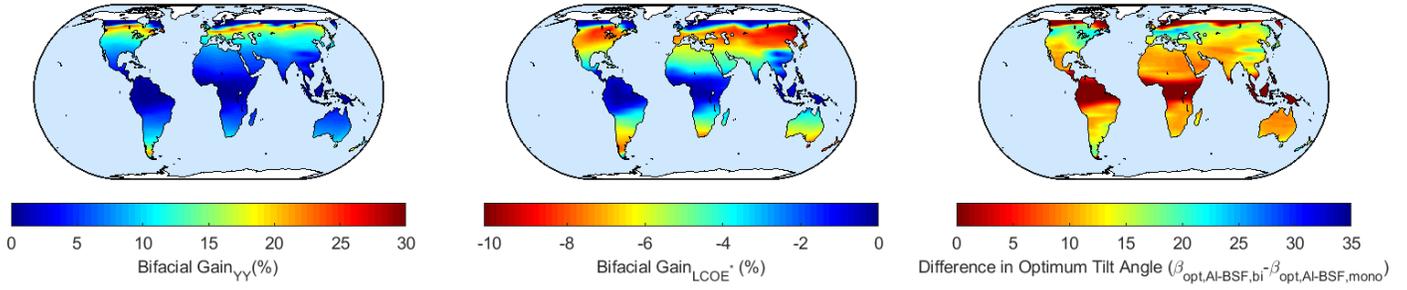

(SHJ (Bifacial) − Al-BSF (Monofacial)) / Al-BSF (Monofacial) for $\eta(T)$

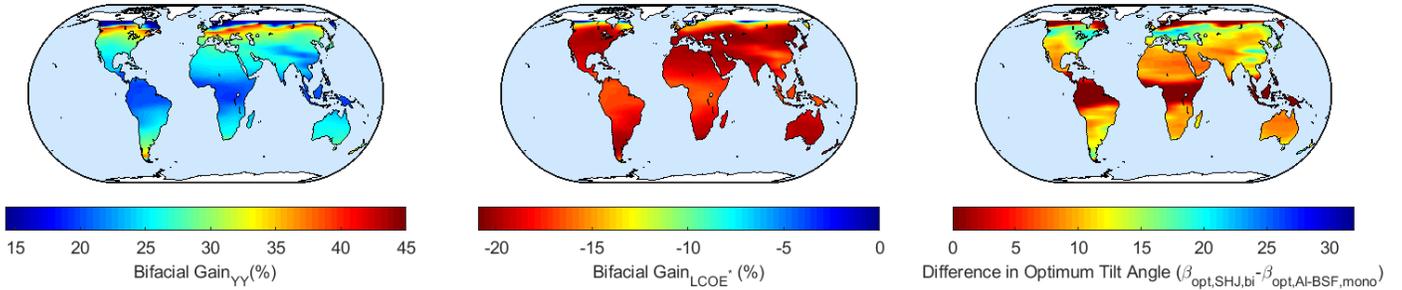

Fig. 7: Global maps comparing bifacial gain in yearly energy yield, $LCOE^*$, and difference in optimum array tilt angle for Al-BSF monofacial farms ($TC = 0.41\%/°C$) vs. (Row 1) Al-BSF and (Row 2) Silicon Heterojunction bifacial farms ($TC = 0.26\%/°C$). Note that $R_A = 0.5$ and $M_L = 15$.

average temperatures, show an improvement in efficiency, with respect to $\eta_{STC}$, leading to higher energy yield.

This general trend is quantitatively shown in Fig. 6 (row 1) for monofacial farms: globally, we observe −7% to +12% of change in energy yield with respect to energy yield estimated using constant efficiency. Fig. 6 (row 2) shows a lower magnitude of change (−7% to +10%) for bifacial farms. Here we assume the same temperature coefficient ($TC = 0.41\%/°C$) for monofacial and bifacial modules (rows 1 and 2 in Fig. 6) with the same Al-BSF technologies. Since the amount of light collected by the bifacial panels ($P_{POA}$) is higher than the monofacial panels, the change in cell/module temperature, from Eq. (13) in Table II, is higher for bifacial modules leading to higher change in efficiency. However, the normalized output (efficiency) for bifacial modules is higher than the monofacial modules (as shown by the y-intercepts in Fig. 1(a)). Overall, the combination of higher normalized output and higher degradation of efficiency for bifacial modules leads to an approximately similar percentage change in energy yield between temperature-dependent vs. constant efficiencies. Moreover, Fig. 6 (row 3) displays the same quantities for bifacial silicon heterojunction technology which has a lower $TC$ (= 0.26%/°C). Lower $TC$ leads to lesser degradation of efficiency with temperature and eventually the range of change in $YY$ (−5% to + 5% ) is lower as compared to Al-BSF ($TC = 0.41\%/°C$).

Fig. 6 also displays the trend in relative percentage change between $LCOE^*$ estimated using temperature-dependent efficiency with respect to $LCOE^*$ calculated using constant efficiency. Since $LCOE^*$ is inversely proportional to the yearly energy yield ($YY$), hence, the trends are reversed as compared to energy yield. Colder places show decrease in $LCOE^*$ by

∼8%, whereas places with hotter climates show a ∼8% increase for bifacial farms. The change (−8% to + 8%) in $LCOE^*$ after temperature correction is significant and cannot be ignored while estimating the levelized cost of energy of a solar PV farm.

The design of solar modules in a farm (represented by the optimum tilt angle and the associated pitch for no mutual shading) does not change significantly. Fig. 6 (maps on the right) shows the difference in the optimum tilt angle when we consider $\eta_T$ vs. $\eta_0$. This difference varies from −3° to + 3° around the globe. Hence, for all practical purposes, the design optimization presented in [15] does not require adjustment even after temperature-dependent efficiencies are included. However, an accurate estimation of $YY$ and $LCOE^*$ is achieved using a more accurate temperature-dependent efficiency model.

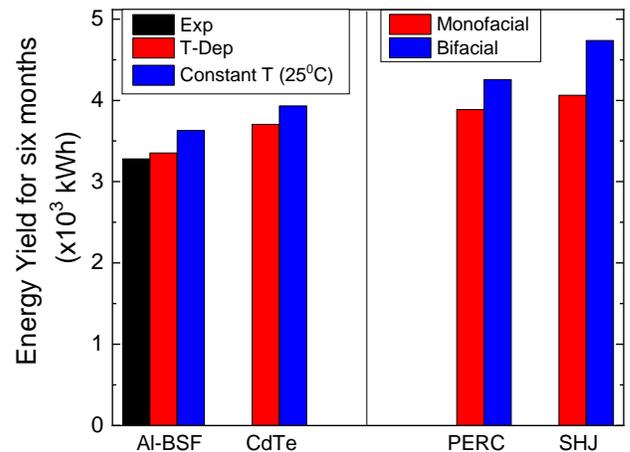

Fig. 8: Energy yield estimated for different technologies illustrating the temperature effect (left) and monofacial and bifacial effect (right).

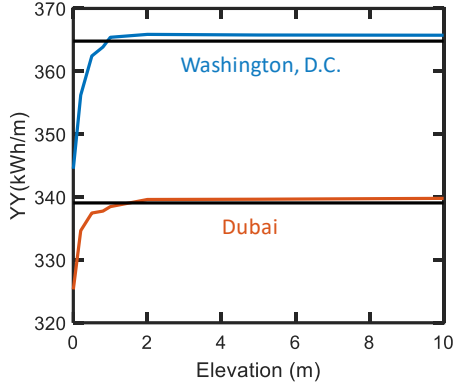

Fig. 9: Temperature – dependent yearly energy yield of the farm (with $p = 2m$ and $h = 1m$) saturates with elevation of panels above the ground. Here we show the simulation for Washington D.C., as an example.

### 3.3 Technology-choice and Geographical location determine the Temperature-dependent bifacial gain (Fig. 7)

Since it has been established that the temperature-dependent model is essential for correct estimation of $YY$ and $LCOE^*$, we will now explain the bifacial gain. We will compare the Silicon Heterojunction (SHJ) ($TC = 0.26\%/°C$) and Aluminum Back Surface Field (Al-BSF) ($TC = 0.41\%/°C$) bifacial panel performance to that of Al-BSF monofacial panels ($TC = 0.41\%/°C$) bifacial panel performance to that of Al-BSF monofacial panels ($TC = 0.41\%/°C$). The global trends in bifacial gains for $YY$ and $LCOE^*$ for constant efficiency calculations can be found in [15]. Fig. 7 displays the same parameters for temperature-dependent Al-BSF (bifacial) (row 1) and SHJ (bifacial) (row 2) versus Al-BSF monofacial farm calculations. We observe that the bifacial gain for yearly yield of Al-BSF is $0-20\%$ for latitudes $< 50°$ and reaches up to 30% for even higher latitudes. The reduction in $LCOE^*$ for bifacial over monofacial is $0-10\%$ around the world. These percentages are higher than that estimated $(2-12\%)$ for temperature-independent (constant) efficiency in Ref. [15]. Whereas, calculations comparing SHJ (bifacial) and Al-BSF (monofacial) show a much higher bifacial gain for yearly yield $(12-45\%)$ and reduction in $LCOE^*$ for bifacial over monofacial parameters $(15-25\%)$ around the world. This is due to higher temperature coefficient for Al-BSF as compared to SHJ panels leading to larger performance degradation of Al-BSF panels and lower bifacial gain. The relative gain simply reflects the fact that the efficiency of bifacial PV degrades less than monofacial PV at a given ambient temperature.

Recall that in Fig. 1, we conceptually compared monofacial and bifacial farms with different efficiencies and temperature coefficients. The observations shown here clearly align with our hypothesis of enhanced temperature-dependent performance for bifacial panels with higher $\eta_{STC}(SHJ) = 19.7\%$ and lower temperature coefficient ($TC(SHJ) = 0.26\%/°C$) as compared to monofacial panels with lower $\eta_{STC}(Al - BSF) = 16.8\%$ and higher temperature coefficient (($TC(Al - BSF) = 0.41\%/°C$).

Clearly, temperature coefficient ($TC$) is an important parameter for accurate efficiency and energy estimation and it varies for different PV technologies. Although bifacial modules collect more light, most bifacial technologies have a lower $TC$ than monofacial ones leading to further enhanced bifacial gain. Furthermore, bifacial technologies do not absorb sub-bandgap (IR) radiation leading to lower heating. Fig. 8 portrays this trend for panels deployed at Sandia National Labs, Albuquerque, and validates our hypothesis. Since temperature coefficient of Al-BSF ($TC = 0.41\%/°C$) is higher than CdTe ($TC = 0.32\%/°C$), the energy yield for former is lower. Further, for both PERC and SHJ, T-dependent estimation shows higher energy yield for bifacial as compared to monofacial due to lower IR absorption. We saw similar results for the global maps in Fig. 7. Thus, we explored the change in energy yield for bifacial farms and monofacial farms with different $TC$.

### 3.4 Temperature-Dependent Energy Yield of Elevated Farms: Self-Heating vs. Improved Light Collection

The energy yield increases with elevation due to improved collection of albedo light at the back face of the panel. Increased light collection increases the temperature, suggesting the need for a temperature-aware model to find optimum elevation and a corresponding bifacial gain. Next, we explore the effect of elevation on temperature-dependent energy yield.

Ref. [39] shows that the energy yield of module starts to saturate at after certain elevation. We observe a similar trend in the farm-level, i.e., yearly energy yield saturates after a certain elevation, as shown in Fig. 9. For Washington D.C., USA, $YY$ increases from 345 kWh/m for ground mounted panel-array ($E = 0$) and saturates to 365 kWh/m at high elevation of panels ($E = 5m$). At elevations $E_0 \sim 1.5\ m$, 95% of this saturated yield can be achieved (for $h = 1m$). Such elevation threshold, where 95% of the maximum/saturated yield would be attained is location specific. For example, in Jeddah, KSA, this elevation threshold is $E_0 = 2m$. We observed that this saturating trend is observed globally. Moreover, the light collection ($P_{POA}$) on the

(SHJ,Bi (E=2m) – SHJ,Bi (E=0m)) / SHJ,Bi (E=0m)

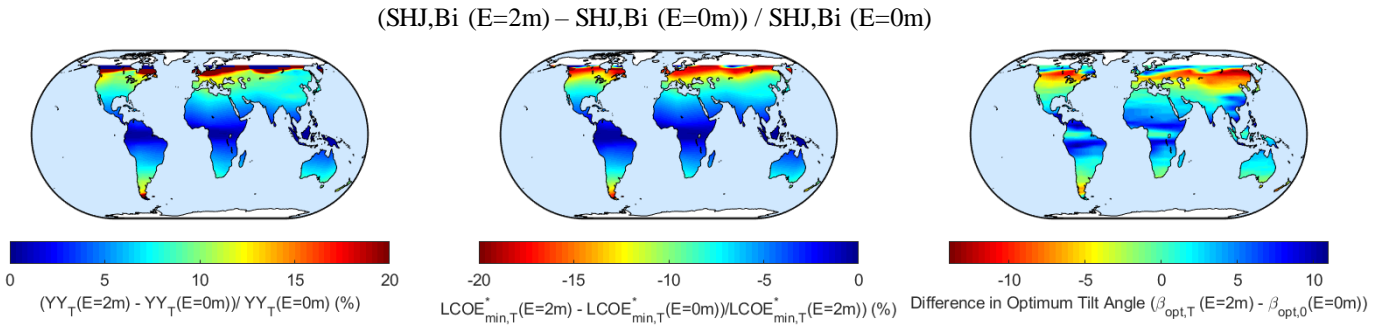

Fig. 10: Global trends in change in yearly energy yield of an elevated bifacial farm ($E = 2m$) and an unelevated bifacial farm ($E = 0m$).

panels increases with elevation, leading to a rise in the temperature-dependent degradation of efficiency (see Eq. 12-14 temp. models). Thus, on one hand elevation increases light collection which enhances energy yield. On the other hand, increasing light collection simultaneously increases the temperature and decreases the efficiency, eventually decreasing the energy yield. Overall, it is observed that the energy yield increases until a threshold elevation ($E_0$) and then saturates to a maximum value.

Fig. 10 shows the percentage change in yearly energy yield and $LCOE^*$ between an elevated bifacial farm (with $E = 2m$) and previously estimated unelevated bifacial farm ($E = 0m$). We chose $E = 2m$ since it is beyond the elevation threshold $E_0$ for energy yield saturation around the world. Evidently, the energy yield improves by $\sim 1 - 20\%$ around the globe with a monotonically increasing trend with latitude. Consequently, $LCOE^*$ decreases by the same amount with a similar trend worldwide. Interestingly, the optimum tilt angle varies from $-10\%$ to $+10\%$ depending on the latitude and the fraction of diffuse light illuminating the bifacial panels. Thus, elevating the farms by $1 - 2m$ enhances the energy yield and reduces $LCOE^*$ (assuming negligible increase in costs). In practice, elevating the modules to $E = 2m$ incurs higher costs due to at least two factors: (1) higher wind loads mean more steal and deeper foundations; (2) installing at that height increases labor costs due to the need for ladders and other work aids. To incorporate this increased cost, a higher value of the cost ratio ($M_L$) in Eq. (3) should be used. Consequently, the optimum value of elevation would be lower, $1m < E_0 < 2m$. A detailed LCOE-dependent analysis of the elevated bifacial farm design will be a part of a future analysis.

## 4. Summary and Conclusion

In this paper, we have analyzed the effects of temperature-dependent efficiency degradation on the energy yield and LCOE of monofacial and bifacial solar PV farms. Our approach involved combining an irradiance model, an updated light collection model for elevated farms, and temperature-dependent efficiency models to arrive at the final energy output of a solar farm. The light collection and temperature estimation models had to be solved self-consistently in order to arrive at the practical and more accurate efficiency for a particular location. We applied these models for locations around the world to deliver the global maps which quantify the percentage change in energy yield and $LCOE^*$ between temperature-dependent and constant temperature calculations, while presenting general global trends.

Our analysis leads to the following key **conclusions**:

- The generalized Alam-Sun thermal model allows us to account for sub-bandgap absorption. The effect is significant: Almost, 90 kWh more energy can be generated from 12 bifacial-module string for a period of 6 months due to the reduced self-heating associated with the transmittance of sub-band irradiance.
- A comparison between energy yield and LCOE for temperature-dependent efficiency ($\eta(T)$) and temperature-

independent efficiency ($\eta_{STC}$) conveys a percentage change of $-7\%$ (Al-BSF), $-5\%$ (SHJ) for locations close to the equator ($|\text{Latitude}| < 30°$) and $+12\%$ (Al-BSF), $+5\%$ (SHJ) for locations close to the poles ($|\text{Latitude}| > 30°$).
- Bifacial gain for SHJ (bifacial) vs. traditionally used Al-BSF (monofacial) with temperature-dependent efficiency conveys a percentage change of $+12\%$ for hotter locations close to the equator ($|\text{Latitude}| < 30°$) and can reach up to $25 - 45\%$ for colder places close to the poles ($|\text{Latitude}| > 30°$). Whereas, bifacial gain for Al-BSF shows a percentage change of $\sim 0 - 30\%$. This presents an incredible opportunity for SHJ bifacial farm deployment.
- The trend in the difference in optimum tilt angle ($\Delta\beta_{opt}$) for bifacial designs is $0° - 30°$ when comparing the same technology (Al-BSF(Bifacial) vs. Al-BSF(Monofacial)) or two different technologies (SHJ(Bifacial) vs. Al-BSF(Monofacial)).
- Elevated farms show two counter-balancing trends, where light collection on the panels ($P_{POA}$) increases leading to increase in temperature and decrease in efficiency. Overall, temperature-dependent elevated farms ($E = 1 - 2m$) outperform unelevated farms in terms of yearly energy yield by $\sim 1 - 20\%$ depending on the latitude.
- Bifacial PV technologies (SHJ) with lower $TC$ and low sub-bandgap (IR) absorption can outperform their monofacial counterparts. The extent of enhancement in performance depends on the bifacial technology used and the geographical location of the farm.

In conclusion, it is important to accurately calculate the energy yield ($YY$) and $LCOE^*$. The bifacial solar farm energy yield using temperature-dependent efficiency fulfills this purpose. Although, the design of the farms in terms of the optimum tilt angle is not affected significantly, but the absolute values of energy yield and LCOE for field-deployed temperature-dependent solar farms differ considerably for several locations around the world. This affects the overall economic evaluation of location-specific solar farms. Since, bifacial panels have a lower temperature coefficient compared to monofacial panels, therefore, they are advantageous for relatively steady energy output due to daily and monthly temperature variations, especially for locations with lower total irradiance. Moreover, using location-specific materials with appropriate temperature coefficients would lead stable outputs and enhanced performance of bifacial solar farms.

Note that temperature-dependent reliability (degradation mechanisms) are not considered in our study. Moving ahead, methods to lower the temperature coefficient of bifacial solar materials/ technologies and mitigation of cell temperature variation through improved packaging would be necessary measures to focus on. These would become a central issue for concentrated solar photovoltaics as well.

ACKNOWLEDGMENT

This work is supported by the National Science Foundation under Grant No. #1724728. R. Ammapet Vijayan gratefully acknowledges the IUSSTF for awarding BASE internship 2019. We are thankful to Josh Stein from Sandia National Labs for providing us the experimental data.

## APPENDIX

### S1: Threshold $P_{POA}$ for $YY_T$ (Bifacial) > $YY_T$ (Monofacial)

The objective is to find the threshold plane of array irradiance $P_{POA}$ for power output of temperature-dependent bifacial panels to be more than that of monofacial panels. We first define efficiency gain ($\eta_g = (\eta_{bi}/\eta_{mo})_{STC}$) as the ratio of normalized output of bifacial module over efficiency of monofacial module at STC. Next, we use $\eta(T_M) = \eta_{STC}(1 - TC(T_M - T_{STC}))$ (Eq. (11) in Table II) and $P_{out} = \eta(T)P_{POA}$ to arrive at the following equation.

$$P_{out} = P_{POA}\eta_{STC}(1 - TC(T_M - T_{STC})). \tag{S1}$$

Assuming $T_M = T_{cell}$ and neglecting the negligible effect of $\eta$ in Eq. (13) of Table II, we get

$$T_{cell} = T_{amb} + kP_{POA} \tag{S2}$$

Here $k$ is a location-specific constant. Now, we take bifacial POA irradiance $(P_{POA})_{bi} = P_{POA}(1 + R_A)$, where $P_{POA}$ is monofacial POA irradiance and $R_A$ is the albedo value. Thus, combining Eqs. S1 and S2, we get a generalized form of output power as shown below.

$$P_{out} = P_{POA}(1 + R_A)$$
$$\times \eta_{STC}(1 - TC(\Delta T_0 + kP_{POA}(1 + R_A))) \tag{S3}$$

where $\Delta T_0 = T_{amb} - T_{STC}$.

For temperature-dependent bifacial output power to be greater than the monofacial counterpart,

$$\frac{(P_{out})_{bi}}{(P_{out})_{mono}} > 1$$

$$\therefore \frac{P_{POA}(1 + R_A)\eta_{bi,STC}(1 - (TC)_{bi}(\Delta T_0 + kP_{POA}(1 + R_A)))}{P_{POA}\eta_{mo,STC}(1 - (TC)_{mono}(\Delta T_0 + kP_{POA}))} \tag{S4}$$

$$> 1$$

Simplifying the above equation gives us,

$$\frac{(1 - (TC)_{bi}(\Delta T_0 + kP_{POA}(1 + R_A)))}{(1 - (TC)_{mono}(\Delta T_0 + kP_{POA}))} > \frac{1}{\eta_g(1 + R_A)} \tag{S5}$$

Eq. S5 is the generalized condition on $P_{POA}$ for bifacial panels to generate higher power output compared to monofacial panels. In special cases, where $(TC)_{bi} = (TC)_{mono} = TC$, $\Delta T_0 = 0$, i.e., $T_{amb} = T_{STC}$, and an extreme value of $R_A = $

1, we can derived the following specialized condition (Eq. S6) mentioned in the introduction of the main text.

$$P_{POA} < \frac{(2\eta_g - 1)}{k \times TC \times (4\eta_g - 1)} \tag{S6}$$

The above equation is physically-intuitive in a sense that bifacial PV would outperform monofacial PV, when the collected plane of array irradiance is lower than a certain threshold value. Moreover, this threshold value should depend only on the above-mentioned parameters which vary with the PV technology ($\eta_g$, $TC$) and the geographical location ($k$, $R_A$).

### S2: Albedo Light Collection

Albedo light collection for elevated panels requires a rigorous estimation of view factors from sky to ground (including masking) i.e., ground pattern as well as collection of albedo light reflected from the ground on the panels. Fig. S1 and S2 display these aspects schematically. Fig. S1 displays the algorithm to calculate the amount light falling on a point on the ground from all parts of the sky that are visible. Note that, there would be only one opening towards the sky for unelevated panels. However, elevated panels allow views of the sky from neighboring periods as well. For each opening (view) of the sky from a point on the ground, we use the view factor formula, $VF = 0.5(\cos(\theta_1) + \cos(\theta_2))$. A product of the total view factor and the intensity of diffuse light gives the total diffuse light falling on a point on the ground. Similarly, for direct light estimation, whenever the direct beam from the Sun falls inside any of the openings (views), we include that into total direct light falling on the point on the ground. A summation of direct and diffuse light falling on that particular point gives us the total illumination on the point on the ground.

Next, we estimate the amount of ground-reflected light collected on the front and rear side of the panel. For unelevated panels, only one period on the ground is visible to the panels, whereas the panels can see multiple periods for elevated panels, see Fig. S2. The above-mentioned view factor (VF) formula is now used to find the amount of light falling from a point on the ground on the panel. An integration over all the points on the ground seen by the panel gives the total light collected on the panel. A product of total intensity of albedo light collected on the panel with efficiency of the module equals the total power output of the panel. An overall integration of total power over time finally leads us to the total energy generated by the panel due to albedo light. A detailed explanation for estimation of albedo light collection for elevated farms can be found in Ref. [40,41].

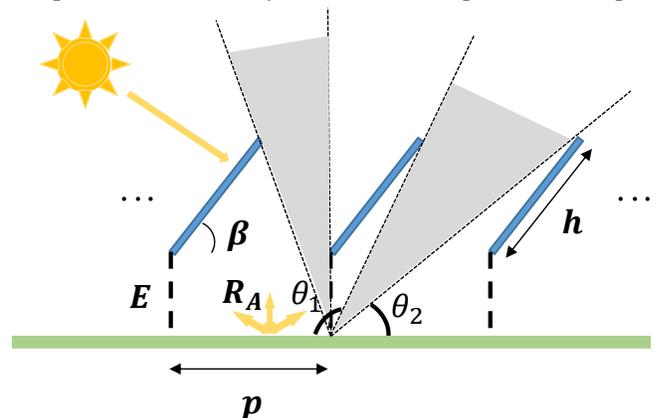

Fig. S1: Ground pattern: illumination at a point on the ground from the visible sky.

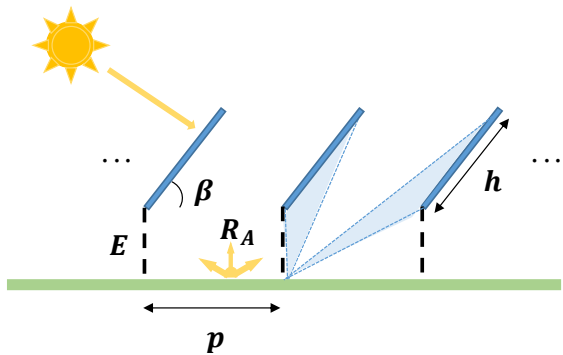

Fig. S2: Light collection: illumination collected by the panel that is reflected from the ground.

## REFERENCES

[1] OCDE & IEA. Market Report Series: Renewables 2017, analysis and forecasts to 2022. Exec Summ 2017:10. doi:10.1073?pnas.0603395103.

[2] Chu S, Majumdar A. Opportunities and challenges for a sustainable energy future. Nature 2012;488:294–303. doi:10.1038/nature11475.

[3] Chu S, Cui Y, Liu N. The path towards sustainable energy. Nat Mater 2016;16:16–22. doi:10.1038/nmat4834.

[4] Kabir E, Kumar P, Kumar S, Adelodun AA, Kim KH. Solar energy: Potential and future prospects. Renew Sustain Energy Rev 2018;82:894–900. doi:10.1016/j.rser.2017.09.094.

[5] Branker K, Pathak MJM, Pearce JM. A review of solar photovoltaic levelized cost of electricity. Renew Sustain Energy Rev 2011;15:4470–82. doi:10.1016/j.rser.2011.07.104.

[6] Ueckerdt F, Hirth L, Luderer G, Edenhofer O. System LCOE: What are the costs of variable renewables? Energy 2013;63:61–75. doi:10.1016/j.energy.2013.10.072.

[7] EIA. Levelized Cost and Levelized Avoided Cost of New Generation Resources in the Annual Energy Outlook 2018 2018:1–20.

[8] Tyagi V V., Rahim NAA, Rahim NA, Selvaraj JAL. Progress in solar PV technology: Research and achievement. Renew Sustain Energy Rev 2013;20:443–61. doi:10.1016/j.rser.2012.09.028.

[9] Green MA, Hishikawa Y, Dunlop ED, Levi DH, Hohl-Ebinger J, Ho-Baillie AWY. Solar cell efficiency tables (version 51). Prog Photovoltaics Res Appl 2018;26:3–12. doi:10.1002/pip.2978.

[10] Chang NL, Ho-Baillie AWY, Vak D, Gao M, Green MA, Egan RJ. Manufacturing cost and market potential analysis of demonstrated roll-to-roll perovskite photovoltaic cell processes. Sol Energy Mater Sol Cells 2018;174:314–24. doi:10.1016/j.solmat.2017.08.038.

[11] sunshot-photovoltaic-manufacturing-initiative @ www.energy.gov n.d. https://www.energy.gov/eere/solar/sunshot-photovoltaic-manufacturing-initiative.

[12] Fu R, Chung D, Lowder T, Feldman D, Ardani K, Fu R, et al. U.S. Solar Photovoltaic System Cost Benchmark : Q1 2017 U.S. Nrel 2017:1–66. doi:10.2172/1390776.

[13] Guerrero-Lemus R, Vega R, Kim T, Kimm A, Shephard LE. Bifacial solar photovoltaics - A technology review. Renew Sustain Energy Rev 2016;60:1533–49. doi:10.1016/j.rser.2016.03.041.

[14] International Technology Roadmap for Photovoltaic—Results 2017 including maturity report 2018. 2018. doi:http://www.itrs.net/Links/2013ITRS/2013Chapters/2013Litho.pdf.

[15] Patel MT, Khan MR, Sun X, Alam MA. A worldwide cost-based design and optimization of tilted bifacial solar farms. Appl Energy 2019;247:467–79. doi:10.1016/j.apenergy.2019.03.150.

[16] Stein JS, Riley D, Deline C, Toor F. Bifacial Solar Photovoltaic Systems : A promising advance in solar performance with interesting challenges 2017.

[17] Alam MA, Khan MR. Thermodynamic efficiency limits of classical and bifacial multi-junction tandem solar cells: An analytical approach. Appl Phys Lett 2016;109. doi:10.1063/1.4966137.

[18] Dubey R, Batra P, Chattopadhyay S, Kottantharayil A, Arora BM, Narasimhan KL, et al. Measurement of temperature coefficient of photovoltaic modules in field and comparison with laboratory measurements. 2015 IEEE 42nd Photovolt Spec Conf PVSC 2015 2015:1–5. doi:10.1109/PVSC.2015.7355852.

[19] King DL, Kratochvil JA, Boyson WE. Temperature Coefficient. Encycl Comput Neurosci 2015:2951–2951. doi:10.1007/978-1-4614-6675-8_100603.

[20] Zhao J, Wang A, Robinson SJ, Green MA. Reduced temperature coefficients for recent high-performance silicon solar cells. Prog Photovoltaics Res Appl 1994;2:221–5. doi:10.1002/pip.4670020305.

[21] Green MA. Solar cells: Operating principles, technology, and system applications. Englewood Cliffs, NJ, Prentice-Hall, Inc.; 1982.

[22] Huebner A, Aberle AG, Hezel R. Temperature behavior of monofacial and bifacial silicon solar cells. Conf Rec IEEE Photovolt Spec Conf 1997:223–6. doi:10.1109/pvsc.1997.654069.

[23] Fan JCC. THEORETICAL TEMPERATURE DEPENDENCE OF SOLAR CELL PARAMETERS AEg mq 3 A k T ] + A OEg AkT3Jsc. Energy 1986;17:309–15.

[24] Lopez-Garcia J, Pavanello D, Sample T. Analysis of temperature coefficients of bifacial crystalline silicon pv modules. IEEE J Photovoltaics 2018;8:960–8. doi:10.1109/JPHOTOV.2018.2834625.

[25] Rodríguez-Gallegos CD, Bieri M, Gandhi O, Singh JP, Reindl T, Panda SK. Monofacial vs bifacial Si-based PV modules: Which one is more cost-effective? Sol Energy 2018;176:412–38.

doi:10.1016/j.solener.2018.10.012.

[26]  Gu W, Ma T, Li M, Shen L, Zhang Y. A coupled optical-electrical-thermal model of the bifacial photovoltaic module. Appl Energy 2020;258:114075. doi:10.1016/j.apenergy.2019.114075.

[27]  Adeh EH, Good SP, Calaf M, Higgins CW. Solar PV Power Potential is Greatest Over Croplands. Sci Rep 2019;9:1–6. doi:10.1038/s41598-019-47803-3.

[28]  Lamers MWPE, Özkalay E, Gali RSR, Janssen GJM, Weeber AW, Romijn IG, et al. Temperature effects of bifacial modules: Hotter or cooler? Sol Energy Mater Sol Cells 2018;185:192–7. doi:10.1016/j.solmat.2018.05.033.

[29]  Khan MR, Hanna A, Sun X, Alam MA. Vertical bifacial solar farms: Physics, design, and global optimization. Appl Energy 2017;206:240–8. doi:10.1016/j.apenergy.2017.08.042.

[30]  Khan MR, Sakr E, Sun X, Bermel P, Alam MA. Ground sculpting to enhance energy yield of vertical bifacial solar farms. Appl Energy 2019;241:592–8. doi:10.1016/j.apenergy.2019.01.168.

[31]  Luque A, Steven H. Handbook of Photovoltaic Science and Engineering. 2nd ed. Wiley; 2011.

[32]  Tillmann P, Jäger K, Becker C. Minimising the Levelised Cost of Electricity for Bifacial Solar Panel Arrays using Bayesian Optimisation 2019:254–64. doi:10.1039/c9se00750d.

[33]  Haurwitz B. Insolation in Relation To Type. J Meteorol 1946;3:123–4. doi:10.1175/1520-0469(1946)003<0123:IIRTCT>2.0.CO;2.

[34]  Haurwitz B. Insolation in Relation To Cloud Amount. Mon Weather Rev 1954;82:317–9. doi:10.1175/1520-0493(1954)082<0317:IIRTCA>2.0.CO;2.

[35]  POWER. Surface meteorology and solar energy: a renewable energy resource web site (release 6.0); 2017.<https://eosweb.larc.nasa.gov/cgi-bin/sse/sse.cgi? > n.d.

[36]  Duffie JA BW. Solar engineering of thermal processes. 4th ed. Wiley; n.d.

[37]  Alam MA, Khan MR. Shockley – Queisser triangle predicts the thermodynamic efficiency limits of arbitrarily complex multijunction bifacial solar cells 2019:1–6. doi:10.1073/pnas.1910745116.

[38]  Barron-Gafford GA, Minor RL, Allen NA, Cronin AD, Brooks AE, Pavao-Zuckerman MA. The photovoltaic heat island effect: Larger solar power plants increase local temperatures. Sci Rep 2016;6:1–7. doi:10.1038/srep35070.

[39]  Sun X, Khan MR, Deline C, Alam MA. Optimization and performance of bifacial solar modules: A global perspective. Appl Energy 2018;212:1601–10. doi:10.1016/j.apenergy.2017.12.041.

[40]  Younas R, Imran H, Riaz MH, Butt NZ. Agrivoltaic Farm Design: Vertical Bifacial vs. Tilted Monofacial Photovoltaic Panels 2019:1–29.

[41]  NREL. Bifacial Radiance Documentation n.d. https://bifacial-radiance.readthedocs.io/en/latest/.

Table III. Glossary of the symbols used in this paper

| Parameters | Definition |
| --- | --- |
| TC | Temperature Coefficient (%/°C) |
| $T_a$ | Ambient Temperature (°C) |
| $T_M$ | Module Temperature (°C) |
| $P_{in}/P_{out}$ | Input Optical Power Collected by the Panel / Output Electrical Power (W/m$^2$) |
| $\eta$ | Power conversion efficiency (%) |
| I/Irradiance | Input optical power over the plane of array (W/m$^2$) |
| a,b | Fitting parameters in the King's model (no unit, s/m) |
| WS | Wind Speed (m/s) |
| $c_T$ | Correction term used for the daily average temperature data |
| $\tau$ | Transmittance of glazing (no unit) |
| $\alpha$ | Absorbed fraction (no unit) |
| $u_L$ | Heat loss coefficient (W/m$^2$K) |
| $\gamma$ | Coefficient of sub-band power contribution to heating (No unit) |
| $\theta_Z$ | Zenith Angle (*degrees*) |
| $\theta_F$ | Angle of incidence at the front face of the panel |
| Pitch (p) | Row-to-row distance between the bottom edges of consecutive arrays (m) |
| Height (h) | Height of the panel |
| E | Elevation |
| $\beta$ | Tilt angle |
| $\gamma_A$ | Azimuth angle from measure from the North |
| $R_A$ | Albedo |
| C/ℂ | Cost / Cost per unit meter |
| M | Number of rows/arrays of modules |
| Z | Number of modules in an array |
| d | Yearly degradation rate in energy conversion |
| Y | Lifetime of a farm (in years) |
| YY | Yearly Yield |
| r | Discount rate |
| F | Fitting parameter in Alam-Sun model which effectively accounts for module assembly and its related effects (like transmittance, glazing, heat loss coefficient) |

| Sub/Super-scripts | Definition |
| --- | --- |
| a | Ambient |
| Alb | Albedo |
| bos | Balance of system |
| bot | Back side |
| DHI | Diffuse Horizontal Irradiance |
| diff | Diffuse |
| dir | Direct |
| dl | Differential element along the height of a panel |
| DNI | Direct Normal Irradiance |
| F | Front |
| f | Fixed cost |
| Farm | PV Farm |
| GHI | Global Horizontal Irradiance |
| Gnd | Ground |
| in | Input |
| l/L | Land |
| M, cell | cell/module |
| om | Operation and maintenance |
| Panel | PV panel |
| PV | Photovoltaic |
| rv | Residual value |
| STC | Standard testing condition |
| sys | System |
| top | top side |
| Total | Total |
| Z | Zenith |